\documentclass[aps,prd,twocolumn,showpacs,preprintnumbers,amssymb,nofootinbib]{revtex4}
\usepackage{amssymb}

\def\be{\begin{equation}}
\def\ee{\end{equation}}
\def\bes{\begin{eqnarray}}
\def\ees{\end{eqnarray}}

\def\half{{\textstyle{1\over2}}}

\catcode`\@=11
\def\@citex[#1]#2{%
\if@filesw \immediate \write \@auxout {\string \citation {#2}}\fi
\@tempcntb\m@ne \let\@h@ld\relax \def\@citea{}%
\@cite{%
  \@for \@citeb:=#2\do {%
    \@ifundefined {b@\@citeb}%
      {\@h@ld\@citea\@tempcntb\m@ne{\bf ?}%
      \@warning {Citation `\@citeb ' on page \thepage \space undefined}}%
      {\@tempcnta\@tempcntb \advance\@tempcnta\@ne%
      \@tempcntb\number\csname b@\@citeb \endcsname \relax%
      \ifnum\@tempcnta=\@tempcntb 
        \ifx\@h@ld\relax%
          \edef \@h@ld{\@citea\csname b@\@citeb\endcsname}%
        \else%
          \edef\@h@ld{\ifmmode{-}\else--\fi\csname b@\@citeb\endcsname}%
        \fi%
      \else
        \@h@ld\@citea\csname b@\@citeb \endcsname%
        \let\@h@ld\relax%
      \fi}%
    \def\@citea{,\penalty\@highpenalty\,}%
  }\@h@ld
}{#1}}

\def\@citeb#1#2{{[#1]\if@tempswa , #2\fi}}
\def\@citeu#1#2{{$^{#1}$\if@tempswa , #2\fi }}
\def\@citep#1#2{{#1\if@tempswa , #2\fi}}

\def\bcites{         
        \catcode`\@=11
        \let\@cite=\@citeb
        \catcode`\@=12
}

\def\upcites{         
        \catcode`\@=11
        \let\@cite=\@citeu
        \catcode`\@=12
}

\def\plaincites{      
        \catcode`\@=11
        \let\@cite=\@citep
        \catcode`\@=12
}

\newcount\hour
\newcount\minute
\newtoks\amorpm
\hour=\time\divide\hour by 60
\minute=\time{\multiply\hour by 60 \global\advance\minute by-\hour}
\edef\standardtime{{\ifnum\hour<12 \global\amorpm={am}%
        \else\global\amorpm={pm}\advance\hour by-12 \fi
        \ifnum\hour=0 \hour=12 \fi
        \number\hour:\ifnum\minute<10 0\fi\number\minute\the\amorpm}}
\edef\militarytime{\number\hour:\ifnum\minute<10 0\fi\number\minute}

\def\draftlabel#1{{\@bsphack\if@filesw {\let\thepage\relax
   \xdef\@gtempa{\write\@auxout{\string
      \newlabel{#1}{{\@currentlabel}{\thepage}}}}}\@gtempa
   \if@nobreak \ifvmode\nobreak\fi\fi\fi\@esphack}
        \gdef\@eqnlabel{#1}}
\def\@eqnlabel{}
\def\@vacuum{}
\def\marginnote#1{}
\def\draftmarginnote#1{\marginpar{\raggedright\scriptsize\tt#1}}
\def\draft{
        \pagestyle{plain}
        \overfullrule=2pt
        \oddsidemargin -.5truein
        \def\@oddhead{\sl \phantom{\today\quad\militarytime} \hfil
        \smash{\Large\sl DRAFT} \hfil \today\quad\militarytime}
        \let\@evenhead\@oddhead
        \let\label=\draftlabel
        \let\marginnote=\draftmarginnote
        \def\ps@empty{\let\@mkboth\@gobbletwo
        \def\@oddfoot{\hfil \smash{\Large\sl DRAFT} \hfil}
        \let\@evenfoot\@oddhead}
        \def\@eqnnum{(\theequation)\rlap{\kern\marginparsep\tt\@eqnlabel}%
        \global\let\@eqnlabel\@vacuum}  }

\begin{document}
\title{Correspondence principle for a brane in Minkowski space and vector mesons}
\thanks{Research supported by the DoE under grant DE-FG05-91ER40627.
}
\author{\sc George Siopsis}
\preprint{UTHET-05-0101}
\email{siopsis@tennessee.edu}
\affiliation{Department of Physics
and Astronomy,
The University of Tennessee, Knoxville,
TN 37996 - 1200, USA.}

\date{February 2005}
\begin{abstract}
We consider a 3-brane of positive
cosmological constant (de Sitter) in $D$-dimensional Minkowski space. We show that the
Poincar\'e algebra in the bulk yields a $SO(4,2)$ algebra when restricted to the brane.
In the limit of zero cosmological
constant (flat brane), this algebra turns into the conformal algebra on the brane.
We derive a correspondence principle for Minkowski space analogous to the
AdS/CFT correspondence.
We discuss explicitly the cases of scalar and gravitational fields.
For a 3-brane of finite thickness in the transverse directions, we obtain a spectrum for vector gravitational perturbations which correspond to vector mesons.
The spectrum agrees with the one obtained in truncated AdS space by de T\'eramond and Brodsky provided $D=10$ and the bulk mass scale $M$ is of order the geometric mean of the Planck mass ($\bar M$) on the brane and $\Lambda_{QCD}$ ($M \sim (\bar M \Lambda_{QCD})^{1/2} \sim 10^9$~GeV).
\end{abstract}

\maketitle

\section{Introduction}
The AdS/CFT correspondence~\cite{bibMald} is by now well-established, even though it
has yet to be proved.
It has led to novel approaches to gravity and an understanding of the
hierarchy problem, raising the possibility
of observing strong gravitational effects at TeV scales~\cite{bibRS}.
Moreover, it has served as a useful tool in analyzing Yang-Mills theories
non-perturbatively~\cite{bibRev}.
Although the CFT is not phenomenologically relevant, significant progress was
recently made toward extracting information related to experiments, following
the work of Polchinski and Strassler~\cite{bibPolS}.
They showed that by introducing an IR cutoff in AdS space, one reproduces the
hard scattering behavior of QCD by convoluting the string amplitude (which exhibits
soft high energy behavior) with AdS wavefunctions.
Subsequent work showed that this model reproduces known theoretical and experimental
results in hadronic physics pointing perhaps to a gravity dual of QCD~\cite{bib1,bib2,bib3,bib4,bib5,bib6,bib7,bib8,bib9,bib10,bib11,bib12,bib13,bib14,bib15,bib16,bib17,bib18,bib19}.

Another interesting possibility is of the extra dimensions forming a flat space
of large or even infinite volume~\cite{biba1,biba2,biba3}.
In this case, light Kaluza-Klein modes may dominate even at low energies~\cite{bibb1,bibb2,bibb3},
leading to a modification of Newton's Law of gravity at astronomically large
distances~\cite{bibc1,bibc2,bibc3,bibc4,bibc5,bibc6,bibc7,bibc8,bibc9}.
Reproducing the results on hardonic physics obtained in AdS space is a challenge in flat Minkowski space due to the lack of a
holographic principle.

Here we make an attempt at understanding the spectrum of hadronic resonances
using an extension of the Dvali-Gabadadze-Porrati (DGP) model~\cite{bibb1,bibb2}.
We consider a fat 3-brane of thickness $\sim 1/M$ in the transverse directions, where
$M$ is the $D$-dimensional bulk Planck mass.
We solve the Einstein equations for vector gravitational perturbations and compare the spectrum with the masses of vector mesons.
We obtain agreement with similar results in AdS space by de T\'eramond and Brodsky~\cite{bib16} for $D=10$, provided the mass scale $M$ is
\be\label{eqA0} M \sim \sqrt{\bar M\Lambda_{QCD}} \sim 10^9~\mathrm{GeV} \ee
where $\bar M$ is the four-dimensional Planck mass.
Unlike in AdS space, there is no need for an artificial IR cutoff; the finite
width of the brane acts as a natural cutoff leading to normalizable solutions
of the wave equation.
We also discuss the conformal dimensions of the corresponding operators which
create the hadronic resonances.
Expanding on an idea discussed in~\cite{ego}, we obtain a correspondence principle
which yields conformal dimensions in agreement with expectations.

Our discussion is organized as follows. In section~\ref{sec2} we show how
the conformal algebra on a 3-brane in flat Minkowski space may be obtained
from the isometries of the embedding in analogy with AdS space~\cite{bibMald}.
In section~\ref{sec3} we consider the case of a scalar field.
We solve the wave equation and discuss the conformal dimensions of the
corresponding operators on the brane.
In section~\ref{sec4} we consider gravitational perturbations.
We derive the Einstein equations and solve them for vector gravitational
perturbations which decouple.
We obtain the spectrum of normalizable modes and compare it to similar results
in truncated AdS space~\cite{bib16}.
In section~\ref{secC} we present our conclusions.
Pertinent results in AdS space are summarized in Appendix~\ref{secA}.

\section{Conformal algebra}\label{sec2}

The isometries of AdS space form an algebra which is isomorphic to the
conformal algebra on the boundary. This is essential for the existence of
the AdS/CFT correspondence principle. In flat Minkowski space, the generators
of isometries form the Poincar\'e algebra which is not simply related to a
conformal algebra.
To establish a correspondence principle in Minkowski space, we shall expand on an idea discussed in~\cite{ego}.
In order not to clutter the notation, we shall work with a five-dimensional Minkowski space spanned by coordinates $X^A = (X^0, \vec X)$, $\vec X = (X^1,\dots,X^4)$,
and place the brane at the hyperboloid
\be\label{eq00} X^A X_A \equiv -(X^0)^2 + \vec X^2 = R^2 \ee
Extending the discussion to higher dimensions is straightforward; one need simply select a flat five-dimensional hypersurface in which to embed the brane.

The cosmological constant on this de Sitter brane is
\be\label{eqcosl} \lambda = \frac{3}{R^2} \ee
We recover the DGP model (flat brane)~\cite{bibb1} in the limit $\lambda\to 0$ ($R\to\infty$).

Let us parametrize the Minkowski space using coordinates appropriate for
a dS hypersurface, $(u, \tau, \vec\Omega)$, where $\vec\Omega = (\Omega^1,\dots,\Omega^4)$ with $\vec\Omega^2 = 1$, as
\be\label{eq4} X^0 = u\sinh\tau \ , \ \ \ \ \vec X = u\cosh\tau\, \vec\Omega \ee
The 3-brane~(\ref{eq00}) is then at $u=R$ and the induced metric reads
\be\label{eq5a} ds_{\mathrm{brane}}^2
= -R^2 d\tau^2 + R^2  \cosh^2\tau \, d\Omega^2\ee
To map the Poincar\'e generators onto the brane, note that for $u=R$,
the momenta may be written as
\bes\label{eq5} P_0 &=& \frac{i}{R} \cosh\tau \partial_\tau \nonumber\\
\vec P &=& - \frac{i}{R} \sinh\tau \vec\Omega \partial_\tau - \frac{i}{R\cosh\tau} \vec\nabla \ees
and the generators of the Lorentz group become
\bes\label{eq6} M_{0i} &=& -i\Omega_i \partial_\tau +i\tanh\tau \nabla_i\nonumber\\
M_{ij} &=& -i(\Omega_i\nabla_j - \Omega_j\nabla_i)\ees
They are easily seen to form a $SO(4,2)$ algebra.
The quadratic Casimir of this algebra is
\be\label{eqCas} \mathcal{C}_2 = \frac{1}{2} (M_{AB} M^{AB} + R^2 P_A P^A)\ee
The flat limit is obtained in the limit $R\to\infty$ by scaling
\be\label{eq9} \tau \to \tau/R \ , \ \ \ \Omega^i \to \Omega^i /R \ \ (i=1,2,3)\ee
In this limit, the metric on the brane~(\ref{eq5a}) turns into
\be ds_{\mathrm{brane}}^2 \approx dx^\mu dx_\mu = - d\tau^2 + (d\Omega^1)^2 + (d\Omega^2)^2 + (d\Omega^3)^2\ee
i.e., flat Minkowski space spanned by Cartesian coordinates $x^\mu = (\tau, \Omega^1,\Omega^2,\Omega^3)$.

To see the fate of the $SO(4,2)$ algebra,
let us introduce the operators
\bes\label{eq8} \mathcal{M}_{\mu\nu} &=& M_{\mu\nu} \ , \ \ \ \ \mathcal{K}_\mu = 2R^2
\left( P_\mu + \frac{M_{\mu y}}{R} \right) \ , \nonumber\\
\mathcal{P}_\mu &=& \frac{1}{2}
\left( P_\mu - \frac{M_{\mu y}}{R} \right) \ , \ \ \ \ \mathcal{D} = RP_y\ees
In the limit $R\to\infty$, they turn into
\bes \mathcal{M}_{\mu\nu} \approx i(x_\mu \partial_\nu -x_\nu\partial_\mu) \ , \ \ \ \mathcal{P}_\mu \approx i\partial_\mu \ , \nonumber\\
\mathcal{D} \approx i x^\mu \partial_\mu \ , \ \ \ \mathcal{K}_\mu \approx
i (x^2 \partial_\mu - 2x_\mu x\cdot \partial) \ees
where $\partial_\mu = \partial/\partial x^\mu$.
These are the generators of the conformal group.

A primary field $\mathcal{O}_\Delta (x)$ of weight $\Delta$ satisfies
\bes [\mathcal{M}_{\mu\nu} \ , \ \mathcal{O}_\Delta (x) ] &=& i(x_\mu \partial_\nu -x_\nu\partial_\mu)\mathcal{O}_\Delta (x) \nonumber\\
\phantom{a} [\mathcal{P}_\mu \ , \ \mathcal{O}_\Delta (x) ] &=& i\partial_\mu \mathcal{O}_\Delta (x) \nonumber\\
\phantom{a} [\mathcal{D} \ , \ \mathcal{O}_\Delta (x) ] &=& i(x^\mu \partial_\mu - \Delta)
\mathcal{O}_\Delta (x) \nonumber\\
\phantom{a} [\mathcal{K}_\mu \ , \ \mathcal{O}_\Delta (x) ] &=& i (x^2 \partial_\mu - 2x_\mu x\cdot \partial + 2\Delta x_\mu)\mathcal{O}_\Delta (x) \nonumber\\
\ees
The quadratic Casimir~(\ref{eqCas}) on the primary field $\mathcal{O}_\Delta (x)$ gives
\be \mathcal{C}_2  \mathcal{O}_\Delta (x) = \Delta (\Delta -4)  \mathcal{O}_\Delta (x)\ee
and the two-point function is
\be G_\Delta (x)\equiv \langle \mathcal{O}_\Delta (\xi) \mathcal{O}_\Delta (0) \rangle \sim \frac{1}{(x^\mu x_\mu)^\Delta}\ee
Thus, we have shown that the Poincar\'e generators in the embedding are mapped
onto the generators of the conformal group on the brane. To establish this
map, it was necessary to bend the brane by introducing a small (positive)
cosmological constant $\lambda$~(eq.~(\ref{eqcosl})) and then take the flat limit, $\lambda\to 0$.
Next, we turn to the wave equation in the embedding to explicitly realize a
correspondence principle in analogy with the AdS/CFT correspondence.

\section{Scalar field}\label{sec3}

Here we consider a massless scalar field in $D$-dimensional Minkowski space in which a 3-brane
resides, which generalizes the DGP model~\cite{bibb1}.
We solve the wave equation in the bulk. By examining its behavior near
the brane, we realize a correspondence principle for the scalar field.
Inclusion of the brane leads to singular expressions which ought to be regulated.
We do this by giving the brane a finite width in the transverse directions.
We show that this affects the conformal dimensions of the operators corresponding to the solutions of the wave equation.

\subsection{In the bulk}

First, consider a flat 3-brane.
Coordinates in the brane will be denoted by $x^\mu$ ($\mu = 0,1,2,3$); transverse coordinates will be $y^a$ ($a=1,\dots, D-4$).
The brane is assumed to be the hypersurface $\vec y = \vec 0$.

The action for a massless scalar field in the bulk is
\be S_{\mathrm{bulk}} = M^{D-2}\int d^4 x d^{D-4}y \partial_A \Phi \partial^A \Phi
\ee
where $M$ is the $D$-dimensional Planck mass.
After Fourier transforming along the brane, we obtain the $D$-dimensional bulk wave equation
\be\label{eq1} (\nabla_y^2 +p^2) \Phi (p, \vec y) = 0 \ee
To solve this, expand in harmonics,
\be \Phi (p,\vec y) = \sum_{L,\vec m} \Phi_{L\vec m} (p, y) Y_{L\vec m} (\Omega_y)\ee
For the $L$th partial wave (suppressing the $\vec m$ indices) we have to solve the
radial wave equation
\be\label{eqll} \frac{1}{y^{D-5}} (y^{D-5} \Phi_L')' + p^2 \Phi_L - \frac{L(L+D-6)}{y^2}
\Phi_L = 0 \ee
whose well-behaved solution is
\be\label{eq17} \Phi_L (p, y) = \mathcal{A}\ y^{-(D-6)/2} H_\alpha^{(1)} (py) \ \ , \ \ \alpha = L + \frac{D-6}{2} \ee
where we included a normalization constant.
This leads to the Green function
\be\label{eq18} G(X^A, X^{\prime A}) = \int \frac{d^4 p}{(2\pi)^4} e^{ip\cdot (x-x')} \Phi_L^*(p,y)\Phi_L (p,y')\ee
where $X^A = (x^\mu, y^a)$ and similarly for $X^{\prime A}$.
After a Wick rotation, the integral may be calculated.
In the limit $y,y'\to 0$ (approaching the brane), we obtain
\be\label{eq20} G(X^A, X^{\prime A}) \sim \frac{(yy')^{L+2}}{((x-x')^2)^\Delta} \ \ , \ \ \Delta = 2+\alpha = L+ \frac{D-2}{2}\ee
In the case $D=10$, we have $\Delta = L+4$, and the scaling agrees with the AdS result~\cite{bibGKP}.
These solutions correspond to operators carrying $SO(6)$ charge.
We shall show that the scaling dimension is modified when the fluctuations of the brane
in the transverse directions are properly accounted for.

The above result~(\ref{eq20}) may also be derived by projecting the $D$-dimensional Green function onto the $L$th partial wave,
\be\label{eq20a} G(X^A, X^{\prime A}) \sim \int d\Omega\ Y_{L\vec m}(\Omega) \frac{1}{((X-X')^2)^{(D-2)/2}}\ee
where
\be\label{eq21} (X-X')^2 = (x-x')^2 + y^2+y^{\prime 2} - 2yy'\cos\theta\ee
$\theta$ being the angle between the vectors $\vec y$ and $\vec y'$.

For a de Sitter brane, parametrize the coordinates as in~(\ref{eq4}), with
$\vec X = (X^1,\dots,X^4)$.
The position of the brane is then given by
\be u = R \ , \ \ \ X^i = 0 \ \ (i=5,\dots, D-1)\ee
Then the distance of two points $X^A$ and $X^{\prime A}$ on the brane is given by
\be\label{eq23} (X^A - X^{\prime A})^2 = 2R^2 (1+\sinh\tau \sinh\tau'-\cosh\tau \cosh\tau' \vec\Omega\cdot\vec\Omega')\ee
Off the brane, we may use~(\ref{eq21}) where $\vec y$ and $\vec y'$ are small
normal vectors.
Since~(\ref{eq20a}) is still applicable, we conclude ({\em cf.}~eq.~(\ref{eq20}))
\be\label{eq20b} G(X^A, X^{\prime A}) \sim \frac{(yy')^{L+2}}{(1+\sinh\tau \sinh\tau'-\cosh\tau \cosh\tau' \vec\Omega\cdot\vec\Omega')^\Delta} \ee
where we used~(\ref{eq23}).
This expression transforms appropriately under the $SO(4,2)$ algebra generated by~(\ref{eq5}) and (\ref{eq6}).
Eq.~(\ref{eq20b}) reduces to~(\ref{eq20}) in the flat-brane limit (after scaling~(\ref{eq9}) and letting $R\to\infty$, with
$(x-x')^2 = -(\tau-\tau')^2 + (\Omega^i - \Omega^{\prime i})^2$ ($i=1,2,3$)).
\subsection{On the brane}
The above expressions ought to be regulated as we approach the brane in order
to obtain quantities defined {\em on} the brane~\cite{bibb2,bibGab}.
This can be done by including fluctuations of the brane in the transverse directions~\cite{bibfb,egoM,bibDR,bibfb2}.
We shall therefore assume that the brane has a finite transverse spread given by
a uniform density $\sigma_\Lambda (y)$ ($y = |\vec y|$)
where $1/\Lambda$ is the width and
\be\label{eq28} \int d^{D-4} y\ \sigma_\Lambda (y) = 1 \ \ , \ \ \lim_{\Lambda\to\infty}
\sigma_\Lambda = \delta^{D-4} (\vec y) \ee
We shall assume that the brane is flat; including curvature, as above, is
straightforward.
The effects of the brane are governed by the action
\be S_{\mathrm{brane}} = \bar M^2 \int d^4 xd^{D-4} y \sigma_\Lambda (y) \partial_\mu\Phi \partial^\mu\Phi\ee
where $\bar M$ is the four-dimensional Planck mass
and we assume
\be\label{eqass} M \sim\Lambda\ll \bar M\sim 10^{19}~\mathrm{GeV}\ee
The wave eq.~(\ref{eq1}) is modified to
\be\label{eq1m} \left( M^{D-2} (p^2 + \nabla_y^2 ) + \bar M^2 p^2 \sigma_\Lambda (y) \right) \Phi (p, \vec y) = 0 \ee
For an explicit calculation, choose a constant density,
\be\label{eq32} \sigma_\Lambda (y) \sim \omega_{D-4} \Lambda^{D-4} \Theta (1/\Lambda - y) \ee
where $\omega_n = \frac{2\pi^{n/2}}{\Gamma(n/2)}$.

Outside the brane (in the bulk, $y>1/\Lambda$), the wave equation reduces to~eq.~(\ref{eqll}) for the $L$th partial wave.
The solution is given by eq.~(\ref{eq17}).

Inside the brane ($y<1/\Lambda$), the solution can be written as
\be\label{eq33} \Phi_L^\mathrm{brane} (p; y) = \mathcal{B} (y\Lambda)^{-(D-6)/2} J_\alpha (\kappa py) \ \ , \ \ \kappa^2 = 1+\frac{\sigma_\Lambda \bar M^2}{M^{D-2}} \ee
Matching expressions across the boundary ($y=1/\Lambda$), we obtain
\bes \mathcal{A} H_\alpha^{(1)} (p/\Lambda) &=&
\mathcal{B} J_\alpha (\kappa p/\Lambda) \ \ , \nonumber\\
\mathcal{A} H_\alpha^{(1)\prime} (p/\Lambda) &=&
\kappa \mathcal{B} J_\alpha' (\kappa p/\Lambda) \ees
For $p\ll \Lambda \sim M$, these can be written as
\bes - \mathcal{A} \frac{i}{\pi} 2^\alpha \Gamma(\alpha) (p/\Lambda)^{-\alpha} &=&
\mathcal{B} J_\alpha (\kappa p/\Lambda) \ \ , \nonumber\\
\mathcal{A} \frac{i\alpha}{\pi} 2^\alpha \Gamma(\alpha) (p/\Lambda)^{-\alpha-1} &=&
\kappa \mathcal{B} J_\alpha' (\kappa p/\Lambda) \ees
These conditions are compatible if
\be (\kappa p/\Lambda) J_\alpha' (\kappa p/\Lambda) + \alpha J_\alpha (\kappa p/\Lambda) = 0\ee
or, equivalently,
\be\label{eqqc} J_{\alpha -1} (\kappa p/\Lambda) = 0\ee
where we used the Bessel function identity
\be\label{eqbfi} zJ_\alpha' (z) + \alpha J_\alpha (z) = z J_{\alpha -1} (z) \ee
Another identity that will be useful later is
\be\label{eqbfi2} zJ_\alpha' (z) - \alpha J_\alpha (z) = -z J_{\alpha +1} (z) \ee
Thus, we obtain the eigenvalues
\be\label{eqei} p^2 = \frac{\beta_{\alpha -1,k}^2 \Lambda^2}{\kappa^2} \ \ , \ \ \alpha - 1 = L + \frac{D-8}{2}\ee
where $\beta_{\alpha -1,k}$ is the $k$th root of $J_{\alpha -1}$.
For $\Lambda\sim M$, we have
\be p^2 \sim \beta_{\alpha-1,k}^2 \frac{M^4}{\bar M^2} \ee
so $p^2 \ll \Lambda^2 \sim M^2$ on account of $M\ll \bar M$, validating our
approximations (unless we choose a root of the Bessel function of very high order).

To establish a correspondence principle, we ought to determine the conformal
dimension of the operator on the brane which corresponds to these solutions
of the wave equation~(eq.~(\ref{eq33})).
In the momentum regime we are working ($p\ll\Lambda$), the tail of the wavefunction~(\ref{eq17}) in the bulk gives a negligible contribution.
An examination of the scaling behavior of the Green function in the bulk led to
a conformal dimension $\Delta = 2+\alpha$ (eq.~(\ref{eq20})). However, this conclusion did not take into account the effects of the finite width of the brane.
We shall determine the conformal dimension by calculating the Green function
on the brane. To this end, it is desirable to analytically continue the momentum
beyond the discrete spectrum~(\ref{eqei}).
Then the normalization constant $\mathcal{B}$ is fixed by going into the UV
regime ($p^2\gg \kappa^2/\Lambda^2\sim \bar M^2 / M^4$).
Defining the inner product
\be\label{eq42} \langle p|p'\rangle \equiv \half \int_{y<1/\Lambda} d^{D-4} y \Phi_L^\mathrm{brane}(p;y) \Phi_L^\mathrm{brane}(p';y)\ee
for the wavefunctions~(\ref{eq33}) on the brane,
we may approximate it in the UV regime by
\be \langle p|p'\rangle
\approx \frac{\omega_{D-4}}{2}\mathcal{B} (p) \mathcal{B} (p')\int_0^\infty dy\, y\, J_\alpha (\kappa py)
J_\alpha (\kappa p' y) \ee
Using the orthogonality property of Bessel functions,
\be \int_0^\infty dy\, y\, J_\alpha (qy) J_\alpha (q'y) = 2 \delta (q^2-q^{\prime 2}) \ee
we deduce $\langle p|p'\rangle = \delta(p^2-p^{\prime 2})$, provided
\be\mathcal{B}^2 = \omega_{D-4} \kappa^2 \sim \omega_{D-4}^2 \frac{\bar M^2}{M^2}\ee
which is momentum-independent and may be ignored in our subsequent discussion.

For momenta satisfying $p\ll\Lambda$, the inner product of two wavefunctions~(\ref{eq33})
may be computed using standard manipulations of Bessel functions.
Ignoring a momentum-independent overall factor, we obtain
\bes \langle p|p'\rangle &\sim& \frac{\kappa p}{\Lambda\mathcal{C}(p,p')} J_\alpha' (\kappa p/\Lambda)J_\alpha (\kappa p'/\Lambda)\nonumber\\
&-& \frac{\kappa p'}{\Lambda\mathcal{C}(p,p')} J_\alpha' (\kappa p'/\Lambda)J_\alpha (\kappa p/\Lambda)\nonumber\\ \ees
where $\mathcal{C}(p,p') = (\kappa p/\Lambda)^2 - (\kappa p'/\Lambda)^2$.
Using the identity~(\ref{eqbfi}), we may write this as
\bes\label{eq46} \langle p|p'\rangle &\sim& \frac{\kappa p}{\Lambda\mathcal{C}(p,p')} J_{\alpha-1} (\kappa p/\Lambda)J_\alpha (\kappa p'/\Lambda)\nonumber\\ 
&-& \frac{\kappa p'}{\Lambda\mathcal{C}(p,p')} J_{\alpha-1} (\kappa p'/\Lambda)J_\alpha (\kappa p/\Lambda)\nonumber\\ \ees
Evidently, for $p\ne p'$, we have $\langle p|p'\rangle = 0$ for eigenstates corresponding to the discrete spectrum~(\ref{eqei}) (obeying the boundary condition~(\ref{eqqc})).
For the normalization of these states, we may use l'H\^opital's rule on the right-hand side of~(\ref{eq46}).
In the limit $p'\to p$, we obtain
\bes\label{eqlh1} \langle p|p\rangle &\sim& \frac{1}{2} [J_{\alpha-1} (\kappa p/\Lambda)]^2 -\frac{\alpha\Lambda}{\kappa p} J_{\alpha-1} (\kappa p/\Lambda) J_\alpha (\kappa p/\Lambda)\nonumber\\
&+& \frac{1}{2} [J_\alpha (\kappa p/\Lambda)]^2 \ees
which reduces to
\be\label{eqlh} \langle p|p\rangle \sim  \frac{1}{2} [J_\alpha (\kappa p/\Lambda)]^2 \ee
for momenta satisfying the boundary condition~(\ref{eqqc}).

The Green function on the brane is defined by averaging the two-point function~(\ref{eq18}) over the transverse spread of the brane,
\be G^{\mathrm{brane}} (x^\mu , x^{\prime\mu}) \equiv \int d^{D-4}y\, \sigma_\Lambda (y) G(x^\mu, \vec y; x^{\prime\mu}, \vec y)\ee
Using the wavefunctions on the brane~(\ref{eq33}),
we may write this Green function in terms of the inner product~(\ref{eq42}) as
\be G^{\mathrm{brane}} (x^\mu , x^{\prime\mu}) \sim \int \frac{d^4p}{(2\pi)^4}\,
e^{ip\cdot (x-x')} \langle p|p\rangle \ee
In the small momentum (IR) limit, eq.~(\ref{eqlh1}) gives
\be \langle p|p\rangle \sim p^{2(\alpha -1)} \ee
leading to the scaling behavior
\be\label{eq53} G^{\mathrm{brane}} (x^\mu , x^{\prime\mu}) \sim \frac{1}{((x-x')^2)^\Delta} \ , \ \ \ \Delta = 1+\alpha \ee
in disagreement with the bulk result $\Delta = 2+\alpha$ (eq.~(\ref{eq20})).
Notice that the bulk result is also obtained if we na{\"\i}vely extrapolate~(\ref{eqlh}), which is only valid for the discrete spectrum~(\ref{eqei}), to the IR regime.
We have argued that the correct analytic continuation is provided by~(\ref{eqlh1}) and {\em not}~(\ref{eqlh}), the latter being a special case of the former.
The bulk result~(\ref{eq20}) does not lead to the correct correspondence principle for operators on the brane,
unless the observable cannot ``see'' the brane (which may be the case with certain gravitational perturbations).
In general, a calculation {\em on} the brane is needed for the analysis of
conformal behavior.
\section{Gravitational field}\label{sec4}
\subsection{Field equations}
As for a scalar, the action for the gravitational field also consists of a bulk term and a brane term. The bulk term is simply
the $D$-dimensional Einstein-Hilbert action
\be\label{eqgrS1} S_{\mathrm{bulk}} = M^{D-2} \int d^DX \sqrt{-\det g}\ R^{(D)} \ee
where $g$ is the $D$-dimensional metric generating the $D$-dimensional Ricci
scalar $R^{(D)}$.

We obtain the brane contribution to the action by dimensional reduction.
Working with a flat brane at $\vec y = \vec 0$, where $X^A = (x^\mu , \vec y)$,
we may decompose the metric as
\bes ds^2 &\equiv& g_{AB} dx^A dx^B\nonumber\\
&=& \bar g_{\mu\nu} dx^\mu dx^\nu + \hat g_{ab}
(dy^a + A_\mu^a dx^\mu)(dy^b + A_\mu^b dx^\mu)\nonumber\\ \ees
Assuming no dependence on the transverse directions, $\vec y$, the Ricci scalar may be written as~\cite{bibPol}
\bes \mathcal{R}^{(D)} &=& \bar R^{(4)} - \frac{1}{4} \partial_\mu (\ln \det \hat g) \partial^\mu (\ln \det\hat g)\nonumber\\
&-& \frac{1}{4} \hat g^{ab} \hat g^{cd} \partial_\mu \hat g_{ac} \partial^\mu \hat g_{bd} - \frac{1}{4} \hat g_{ab} F_{\mu\nu}^a F^{b\mu\nu} \ees
where $\bar R^{(4)}$ is the four-dimensional Ricci tensor generated by $\bar g_{\mu\nu}$ and
\be F_{\mu\nu}^a = \partial_\mu A_\nu^a - \partial_\nu A_\mu^a \ee
The brane term is
\be\label{eqgrS2} S_{\mathrm{brane}} = \bar M^2 \int d^4 x d^{D-4} y \sqrt{-\det \bar g} \sigma_\Lambda (y) \mathcal{R}^{(D)}
\ee
where we included the effects of fluctuations in transverse directions given
in terms of the profile function $\sigma_\Lambda$ (eq.~(\ref{eq28})).

Let us perturb around the Minkowski flat background
\be g_{AB} = \eta_{AB} + h_{AB} \ee
Varying the action  $S= S_{\mathrm{bulk}} + S_{\mathrm{brane}}$, given by
eqs.~(\ref{eqgrS1}) and (\ref{eqgrS2}), with respect to $h_{AB}$, we obtain the Einstein field equations.
To solve these equations, in general, is challenging, because different gravitational perturbations are coupled to each other~\cite{egoM}.

Here we concentrate on the fluctuations generated by the off-diagonal components of the metric (vector potential),
\be\label{eq60} h_{a\mu} = A_{a\mu} \ee
We obtain the field equation
\bes\label{eqfe0} M^{D-2} (\partial_A\partial^A h_{a\mu} &-& \partial_\mu \partial^A h_{aA}
-\partial_a \partial^A h_{\mu A} + \partial_\mu \partial_a h_A^A)\nonumber\\
&+& \bar M^2 \sigma_\Lambda (y) \partial^\nu F_{a\mu\nu} = 0 \ees
Choosing the harmonic gauge,
\be\label{eqhg} \partial^A h_{AB} = \frac{1}{2} \partial_B h_A^A \ee
eq.~(\ref{eqfe0}) simplifies to
\be\label{eqfe} M^{D-2} \partial_A\partial^A A_\mu^a + \bar M^2 \sigma_\Lambda (y) \partial^\nu F_{\mu\nu}^a = 0 \ee
This field equation is consistent provided
\be \partial_\mu A^{a\mu} = 0 \ee
(as for a massive vector field)
in which case~(\ref{eqfe}) reads
\be\label{eqfe2} M^{D-2} \partial_A\partial^A A_\mu^a + \bar M^2 \sigma_\Lambda (y) \Box A_\mu^a = 0 \ee
Thus we obtained a wave equation for vector gravitational perturbations
decoupled from all other modes.
This equation is of the same form as the scalar wave equation~(\ref{eq1m}).
As we discussed in section~\ref{sec3}, the field with angular momentum $L$ in
the transverse directions ($SO(D-4)$ quantum number) corresponds to an operator
on the brane of conformal dimension
\be\label{eq66} \Delta = 1+\alpha = L + \frac{D-4}{2} \ee
where we used eqs.~(\ref{eq53}) and (\ref{eq17}).
Moreover, the solution of the wave equation~(\ref{eqfe2}) leads to the discrete spectrum~(\ref{eqei}).

\subsection{Vector mesons}

A vector meson of internal angular momentum $L$ has non-vanishing coupling to
the interpolating operator~\cite{bibo1,bibo2,bib16}
\be\label{eqio} \mathcal{O}_{L+3}^\mu = \bar\psi \gamma^\mu D_{\{ \mu_1} \ldots D_{\mu_L\} } \psi \ee
where $D_\mu = \partial_\mu + ig_{YM} A_\mu$, and we have taken the traceless
symmetric part of the tensor.
This operator has dimension $\Delta = L+3$.

Our earlier results on the correspondence between vector perturbations of the metric~(\ref{eq60}) and operators on the brane of
conformal dimension given by~(\ref{eq66}) are not directly applicable to the
interpolating operators~(\ref{eqio}).
This is because the angular momentum quantum number appearing in~(\ref{eq66})
represents rotation in the transverse directions, whereas the operators in~(\ref{eqio}) have spin along the directions of the brane.
In AdS space one obtains agreement in the conformal dimensions of these two
types of operators, corresponding to strings spinning transversely and along the brane, respectively, in the limit of large dimensions with logarithmic corrections~\cite{bibo3}.
A similar argument should apply to our case of flat Minkowski space for $D=10$.
To map the isometries onto the conformal group of the brane, it is necessary to
treat the brane as the flat-space limit of a de Sitter brane, as discussed in section~\ref{sec2}.

Consequently, for $D=10$ and large $L$, the interpolating operator~(\ref{eqio})
corresponds to the $L$th partial wave solution of~(\ref{eqfe2}).
The conformal dimension deduced earlier (eq.~(\ref{eq66})) for $D=10$ is $\Delta = L+3$, in agreement with the QCD result for the operator~(\ref{eqio}).
Moreover, the spectrum deduced from the wave equation~(\ref{eqfe2}), given by eq.~(\ref{eqei}) with $D=10$,
\be\label{eq68} p^2 = \frac{\beta_{L+1,k}^2 \Lambda^2}{\kappa^2} \approx \beta_{L+1,k}^2\, \frac{M^4}{\omega_6\bar M^2}\ee
where we used $\Lambda = M$ and eqs.~(\ref{eq32}) and (\ref{eq33}).
This coincides with the spectrum derived in truncated AdS space~\cite{bib16}
provided
we choose the ten-dimensional mass scale to be
\be M = (\pi^{3/2} \bar M \Lambda_{QCD})^{1/2} \sim 10^9~\mathrm{GeV} \ee
as advertised~(eq.~(\ref{eqA0})).
As was shown in~\cite{bib16}, this spectrum is in good agreement with experimental results~\cite{bibPD}.
It should be noted that in our case, there was no need for an arbitrary IR cutoff.
An effective cutoff was provided by the finite transverse width of the brane,
$1/\Lambda \sim (\bar M\Lambda_{QCD})^{-1/2}$.
We chose a crude profile for the brane (step function~(\ref{eq32})).
The prediction~(\ref{eq68}) of the spectrum may be improved upon by a refinement of
the profile function~(\ref{eq32}).

\section{Conclusions}\label{secC}

In conclusion, we have discussed a correspondence principle for a 3-brane in
$D$-dimensional flat Minkowski space in analogy with the AdS/CFT correspondence~\cite{bibMald}.
Expanding on an idea discussed in~\cite{ego}, we showed that the isometries of the embedding, generating the Poincar\'e group, can be mapped onto a $SO(4,2)$ algebra on the brane if the latter is a de Sitter
hypersurface.
In the flat limit (zero cosmological constant), this $SO(4,2)$ algebra turns
into the conformal algebra on the 3-brane.
We then realized a correspondence principle by considering fields with finite
angular momentum in the transverse directions and showed that
they led to a scaling behavior of the Green function in analogy with the results on the boundary propagator in AdS space~\cite{bibGKP}.
We discussed a subtlety related to the correct analytic continuation of the
spectrum of the $D$-dimensional wave equation to arbitrary momenta which affected the determination of the conformal dimension of the corresponding operator
on the brane.
Finally, we applied our results to the case of vector mesons which corresponded
to vector gravitational perturbations in the bulk.
We obtained a spectrum which was in agreement with AdS results~\cite{bib16}
provided we chose $D=10$ and the ten-dimensional mass scale as in~(\ref{eqA0}).

Extending the discussion to other hadronic resonances entails a more complete solution of the
ten-dimensional Einstein equations. Fermions may also be included by extending
the model to supergravity. Work in this direction is in progress.

\appendix

\section{Review of AdS}\label{secA}

Here we review the pertinent features of AdS space for comparison with our
results in flat Minkowski space.
The wave equation for a massless scalar in AdS$_5\times S^5$ is~\cite{bibGKP}
\be\label{eq0} \partial_z^2\phi - \frac{3}{z} \partial_z \phi + p^2\phi - \frac{L(L+4)}{z^2} \phi = 0\ee
where $p^\mu$ is the four-momentum on the boundary ($z\to 0$) and $L$ is the $S^5$ angular momentum.
The solution is
\be \phi (p; z) = \mathcal{B} z^2 J_\nu (pz) \ \ , \ \ \nu = L+2 \ee
The normalization constant is fixed by imposing the normalization condition
\be (p|p') \equiv \int_0^\infty \frac{dz}{z^3}\, \phi^*(p;z) \phi(p';z) = \delta(p^2 - p^{\prime 2})\ee
leading to a momentum-independent $\mathcal{B}$.

The propagator is given by
\be G (x,z; x', z') = \int \frac{d^4 p}{(2\pi)^4} e^{-ip\cdot (x-x')}
\phi^* (p;z) \phi(p; z') \ee
from which one may deduce the boundary propagator
in the limit
$z,z'\to 0$. We obtain
\bes\label{eqag} G(x,z;x',z') &\sim& (zz')^{\nu+2}\int \frac{d^4 p}{(2\pi)^4} e^{-ip\cdot (x-x')} p^{2\nu}\nonumber\\
&\sim& \frac{(zz')^\Delta}{((x-x')^2)^\Delta} \ \ , \ees
where $\Delta = \nu+2 = L+4$.

We wish to consider wavefunctions which live on the boundary but have a finite
extent in the bulk direction $z$, $0\le z\le z_0$ where $z_0$ is to be identified with the QCD scale ($z_0 = 1/\Lambda_{QCD}$)~\cite{bib16}.
This cutoff $z_0$ breaks conformal invariance and can be thought of as providing a finite width for the brane residing on the boundary of AdS.
Then CFT observables must be defined by averaging over the transverse width
of the brane.
Let us then define the boundary propagator by
\be \mathcal{G} (x,x') \sim \int_0^{z_0} \frac{dz}{z^3} G(x,z;x',z) \ee
The conformal limit is recovered in the IR ($p z_0\ll 1$), because
for small $p$, we have $J_\nu (pz) \sim (pz)^\nu$, therefore
\be\label{eqav} \mathcal{G} (x,x') \sim z^{2(\nu +2)}\int \frac{d^4 p}{(2\pi)^4} e^{-ip\cdot (x-x')} p^{2\nu}
\sim \frac{z^{2\Delta}}{((x-x')^2)^\Delta} \ \ , \ee
in agreement with eq.~(\ref{eqag}).




If we impose the boundary condition
\be\label{eqbcl} J_\nu (pz_0) = 0 \ee
the momentum $p$ is discretized,
\be\label{eqmodi} p^2 = \frac{\beta_{\nu, k}^2}{z_0^2}\ \ , \ \ J_\nu (\beta_{\nu, k}) = 0 \ee
Thus, $p\gtrsim o(1/z_0)$. These momenta no longer satisfy the requirement $pz_0\ll 1$ and we are outside the conformal regime.
Notice that for the discrete spectrum~(\ref{eqmodi}),
\be\label{eq43} \int_0^{z_0} \frac{dz}{z^3} [\phi (p;z)]^2 = \frac{1}{2}
[J_{\nu+1} (pz_0)]^2 \ee
Na{\"\i}vely analytically continuing this to low momenta, we obtain a
behavior of $p^{2(\nu+1)}$ which yields the wrong boundary propagator
({\em cf.}~eq.~(\ref{eqav})).
To do the analytic continuation correctly, we ought to first calculate the
inner product
\be (p|p') = \int_0^{z_0} \frac{dz}{z^3} \phi (p;z) \phi (p';z) \ee
which vanishes for $p,p'$ obeying~(\ref{eqmodi}) with $p\ne p'$, and then
take the limit $p'\to p$.
This yields
\bes\label{eqlh1a} (p|p) &=& \frac{1}{2} [J_\nu (pz_0)]^2 -\frac{\nu}{pz_0} J_\nu (pz_0) J_{\nu+1} (pz_0) \nonumber\\
&+& \frac{1}{2} [J_{\nu+1} (pz_0)]^2\ees
which reduces to~(\ref{eq43}) for momenta satisfying the quantization condition~(\ref{eqbcl}).
Continuing this to low momenta, we obtain $(p|p)\sim p^{2\nu}$ and we recover
eq.~(\ref{eqav}), as expected.

The discussion of the wave equation for a vector in AdS space is similar~\cite{bibV}.

\begin{acknowledgments}
I am indebted to Guy F.~de T\'eramond for fruitful discussions.
\end{acknowledgments}


\end{document}